# Type-II Red Phosphorus: Wavy Packing of Twisted Pentagonal Tubes

Jun-Yeong Yoon,[a,b] Yangjin Lee,[a,b] Dong-Gyu Kim,[a] Dong Gun Oh,[a] Jin Kyun Kim,[c]† Linshuo Guo,[d] Jungcheol Kim,[e] Jeongheon Choe,[a] Kihyun Lee,[a] Hyeonsik Cheong,[e] Chae Un Kim,[c] Young Jai Choi,[a] Yanhang Ma,*[d] Kwanpyo Kim*[a,b]

[a] Department of Physics, Yonsei University, Seoul 03722, Korea.
[b] Center for Nanomedicine, Institute for Basic Science (IBS), Seoul 03722, Korea.
[c] Department of Physics, Ulsan National Institute of Science and Technology (UNIST), Ulsan 44919, Korea.
[d] School of Physical Science and Technology & Shanghai Key Laboratory of High-resolution Electron Microscopy, ShanghaiTech University, Shanghai 201210, China.
[e] Department of Physics, Sogang University, Seoul 04107, Korea.

* E-mail: Y.M. (mayh2@shanghaitech.edu.cn) and K.K. (kpkim@yonsei.ac.kr)

†Present address: Division of Industrial Metrology, Korea Research Institute of Standards and Science, Daejeon 34113, Republic of Korea

**Abstract:** Elemental phosphorus exhibits fascinating structural varieties and versatile properties. The unique nature of phosphorus bonds can lead to the formation of extremely complex structures, and detailed structural information on some phosphorus polymorphs is yet to be investigated. In this study, we investigated an unidentified crystalline phase of phosphorus, type-II red phosphorus (RP), by combining state-of-the-art structural characterization techniques. Electron diffraction tomography, atomic-resolution scanning transmission electron microscopy (STEM), powder X-ray diffraction, and Raman spectroscopy were concurrently used to elucidate the hidden structural motifs and their packing in type-II RP. Electron diffraction tomography, performed using individual crystalline nanowires, was used to identify a triclinic unit cell with volume of 5330 Å$^3$, the largest unit cell for elemental phosphorus crystals up to now, which contains approximately 250 phosphorus atoms. Atomic-resolution STEM imaging, which was performed along different crystal-zone axes, confirmed that the twisted wavy tubular motif is the basic building block of type-II RP. Our study discovered and presented a new variation of building blocks in phosphorus, and it provides insights to clarify the complexities observed in phosphorus as well as other relevant systems.

**Introduction**

Elemental phosphorus forms many different polymorphs.[1-2] With five electrons in its outer shell, the phosphorus structure can be stabilized in various polymorphs by forming bonds with three neighboring atoms in a tetrahedral geometry with one lone pair. The structural diversity of elemental phosphorus and its origins have attracted considerable research interest over the last few decades.[3-7] Moreover, previous theoretical calculations and experiments indicated the presence of numerous unexplored phosphorus polymorphs.[6, 8-17] Considering that various forms of phosphorus crystals have recently emerged as platforms for investigating unique electronic and optical properties in low dimensions,[18-22] the identification of a new structural phase of elemental phosphorus will be of critical importance to better design materials for various applications.

Red phosphorus (RP), which typically exists in an amorphous polymeric (type-I) configuration,[5] can be crystallized into various types of polymorphs.[2] An early study, conducted in 1947, categorized the crystalline phases of RP (types II, III, IV, and V) based on X-ray diffraction (XRD) data and a thermal analysis.[2] The crystal structure of type-V RP (violet or Hittorf's phosphorus) has been identified using single-crystal XRD[23] in 1966 and has recently attracted increasing research interest.[15-16, 24] The crystal structure of type-IV RP (fibrous phosphorus) has been identified in a relatively recent study.[25] Both types-IV and V RP contain a tubular motif [P8]P2[P9]P2[ in their structures.[25] However, the crystal structures of types-II and III are yet to be identified, and this has become a long-standing question regarding phosphorus structures. This may be due to the complex structures of these polymorphs. Type-II RP can be converted from an amorphous phase at lower temperatures than those required for type-IV or type-V syntheses.[2, 26] The synthesis conditions at relatively low temperatures suggest that the structure is complex, sharing the nature of disordered interatomic bond configurations in the amorphous state. Previous studies on type-II materials often focused on the utilization of materials for various applications but lacked detailed structures at the atomistic level owing to incomplete crystal structure information.[27-28]

In this study, we elucidated the crystal structure of type-II RP using state-of-the-art structural characterization techniques involving a combination of three-dimensional (3D) electron diffraction tomography, atomic resolution scanning transmission electron microscopy (STEM), powder X-ray diffraction, and Raman spectroscopy. Using individual crystalline nanowires, 3D electron diffraction tomography was performed to identify the unit cell and possible space groups of type-II RP. Moreover, atomic resolution STEM imaging was performed along different crystal zone axes to confirm the basic building block of type-II RP. Type-II RP has been used in various research disciplines[29-30] and our identification of the type-II RP structure will pave the way for more research aimed at providing a better understanding of phosphorus complexity.

**Results and discussion**

Type-II RP was synthesized using chemical vapor transport (CVT) method. The detailed information on the synthesis can be found in Method section and Supporting Figure S1. The synthesized phosphorus samples exhibited a typical dark red color with a polycrystalline nature, as shown in Figure 1b. Each crystallite was approximately 100 µm in size. After mechanical exfoliation or sonication, the crystals were split to display a nanowire-type morphology (Figure 1c and 1d). The lengths and diameters of the wires were in the ranges of tens of micrometers and a couple of hundred nanometers, respectively. The individual wires were identified as mostly single crystals, as shown by electron diffraction (Figure 1e). Energy dispersive X-ray spectroscopy (EDS) confirmed that the obtained crystal was exclusively composed of phosphorus, without any noticeable presence of other elements (Supporting Figure S2).

Powder XRD and Raman spectroscopy were used as characterization tools to identify the RP phases. We paid particular attention to proper comparison with the previous literature because of the incomplete understanding of type-II RP structures. Figure 1f shows the XRD patterns of the Type-II RP synthesized in our study, as well as those of a previously reported crystalline RP phase. All types of RP have the strongest diffraction peaks around $2\theta = 16°$. Notably, these strong peaks are associated with the inter-tubular distances in type-IV and type-V RP phases.[25] We performed XRD analysis on multiple samples using different instruments (lab-based and synchrotron-source XRD), and all of them yielded consistent results (Supporting Figure S3). The powder XRD data of our sample are consistent with the reference[26, 31] as well as the original report of 1947,[2] indicating that our synthesized sample is type-II RP.

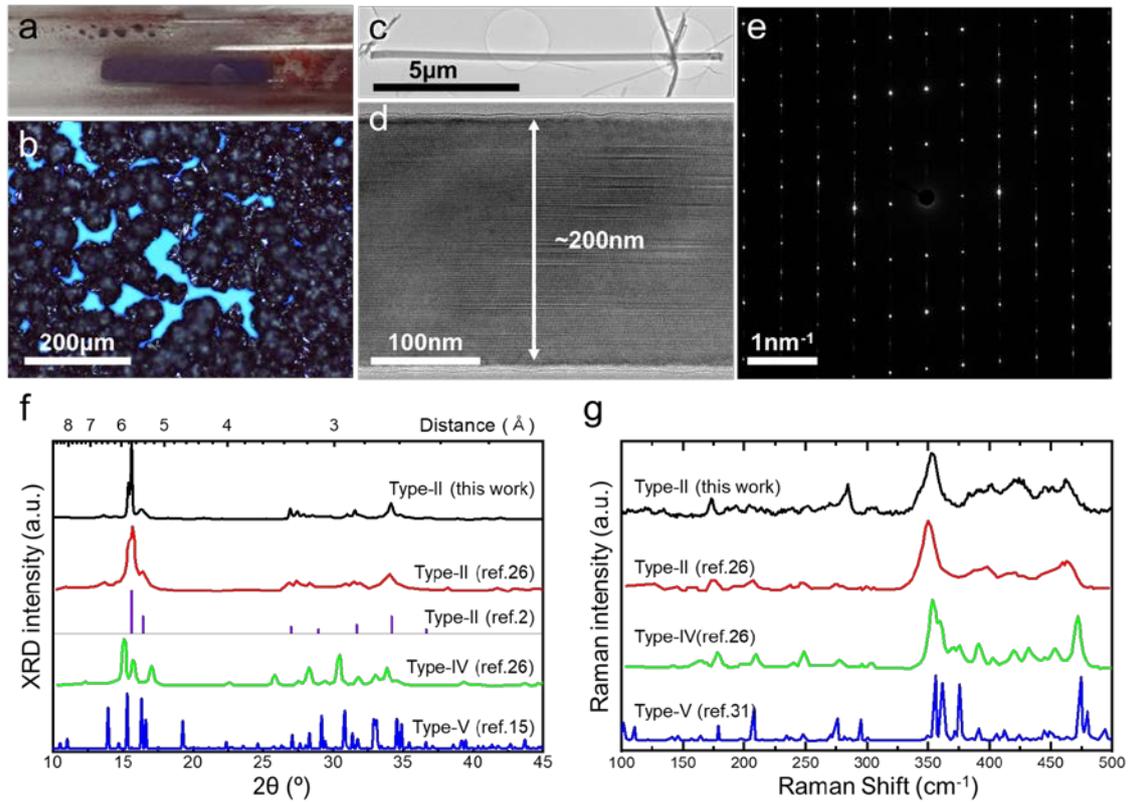

**Figure 1. Synthesis and identification of type-II RP**. (a) Photo of synthesized type-II RP crystals in a quartz ampule. (b) Optical image of type-II RP crystals grown on a $SiO_2$/Si wafer. (c,d) TEM images of the type-II RP nanowire. (e) SAED of the type-II RP wire. (f) Powder XRD data of various types of RP polymorphs. Data from this work, type-II (red), type-IV (green), and type-V (blue). The third entry (purple) is from the original reference, which has categorized various types of RP. (g) Raman spectrum comparison between various types of RP polymorphs.

The Raman spectra of various known RP polymorphs are shown in Figure 1g. The Raman spectra obtained from the samples in the current study are also consistent with previous reports on type-II RP.[26] The polarized Raman spectroscopy was also performed on the individual wire samples (Supporting Figure S4). We observed a strong laser-polarization-direction dependence of the Raman signals with 180-degree rotational symmetry, which also supports that the individual wire is a single crystalline structure. The presence of many Raman peaks also indicates the complex nature of the type-II RP crystal structure. We also performed Raman measurements on well-isolated multiple wire samples and confirmed that the samples had a single phase of type-II RP.

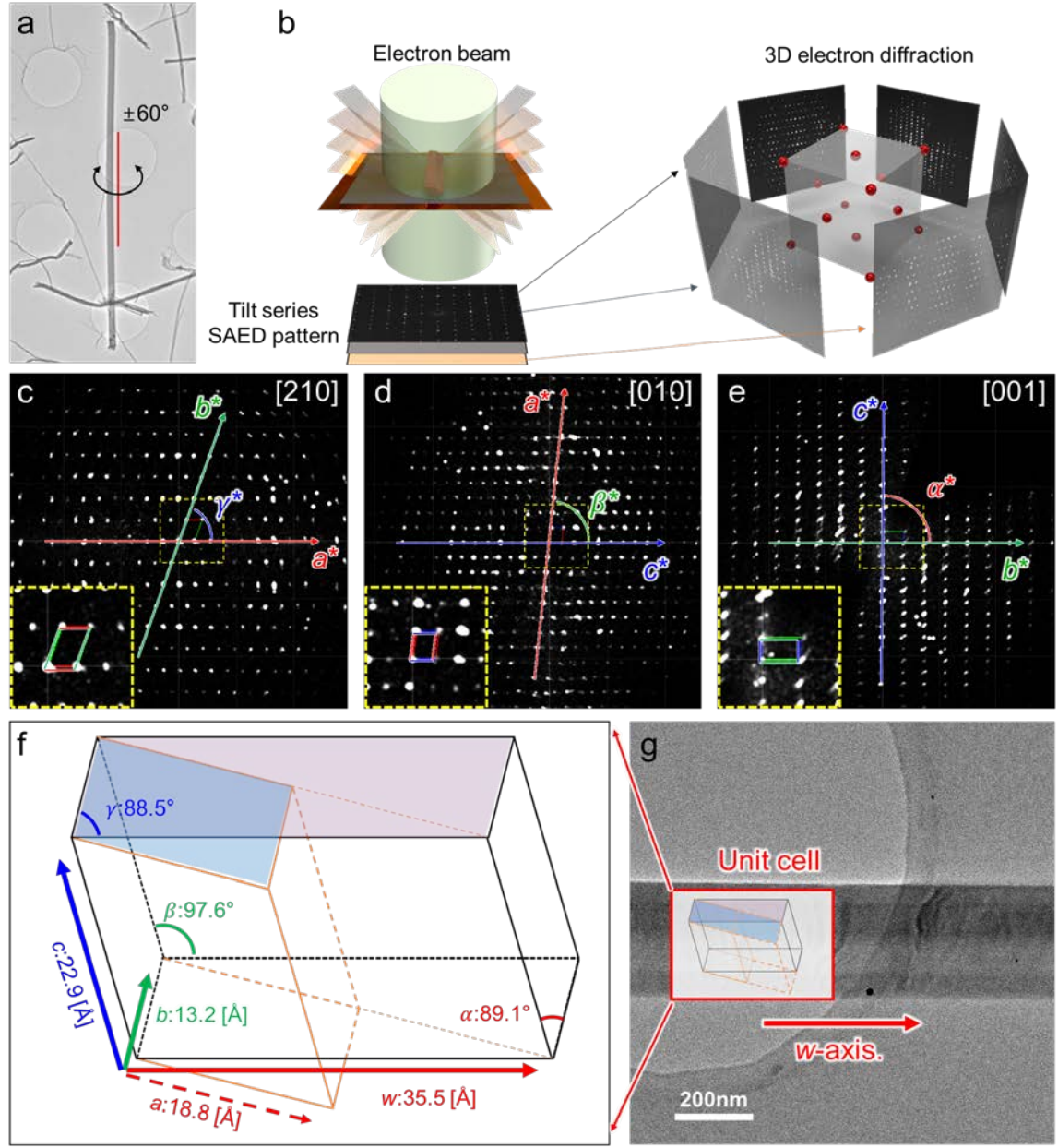

**Figure 2. Three-dimensional electron diffraction and identification of a type-II RP unit cell**. (a) TEM image of a type-II RP nanowire with a tilting condition. (b) Schematic of the tilt series SAED acquisition and reciprocal space reconstruction from the tilt-series SAED data. (c-e) Three-dimensional electron diffraction data projected from different zone axes. Diffraction data along the zone axes of c) [210], d) [010], and e) [001]. The insets show the zoomed-in diffraction near the center peak. (f) Identified primitive unit cell (box with orange outline) and its double-unit cell (box with black outline) of type-II RP. The nanowire axis is parallel to the $w$-axis of the double-unit cell.

We employed 3D electron diffraction to obtain unit cell information on type-II RP. Electron crystallography has recently emerged as an alternative method to conventional X-ray crystallography tools.[32-34] In particular, the sample size required for electron crystallography is much smaller (~10 nm) than that required for single-crystal XRD. This merit is critical for the investigation of type-II RP because large-sized single crystals are unavailable to date. We acquired a tilt series of selected-area electron diffraction (SAED) patterns from individual type-II nanowires via transmission electron microscopy (TEM). Approximately 600 SAED patterns with tilt steps of 0.2° were recorded to reconstruct a 3D reciprocal lattice, as shown in Figure 2a and 2b. Figure 2c-2e show representative three-dimensional electron diffraction data projected along low-index zone axes. From the reconstructed data, we assigned the primitive unit cell of type-II RP for the first time: $a = 18.8$ Å, $b = 13.2$ Å, $c = 22.9$ Å, $\alpha = 89.1°$, $\beta = 97.6°$, and $\gamma = 108.8°$, as shown in Figure 2f. The possible space groups of the identified structures were *P*1 and *P*-1. By combining the identified unit-cell information and real-space imaging of the nanowire, we found that the wire axis was aligned along the [210] direction (Figure 2g). Therefore, for the convenience of analysis, we assigned the wire axis (or *w*-axis) along the [210] direction and the double-unit cell, as shown in Figure 2f. The information on the double-unit cell is given as $a = 35.5$ Å, $b = 13.2$ Å, $c = 22.9$ Å, $\alpha = 89.1°$, $\beta = 97.6°$, and $\gamma = 88.5°$. For the following discussion and indexing of diffraction peaks, we used a double-unit cell.

The electron diffraction tomography data from multiple nanowires displayed identical unit cell information, which confirms that the identified unit cell belonged to type-II RP. The identified type-II RP unit cell was large, containing approximately 250 atoms in the primitive unit cell (or 500 atoms in the double-unit cell), based on the expected atomic density of phosphorus. To the best of our knowledge, the identified type-II RP crystal unit cell is one of the largest unit cells identified among elemental crystals. The identified low-symmetry triclinic unit cell explains the difficulty in synthesizing large single crystals.

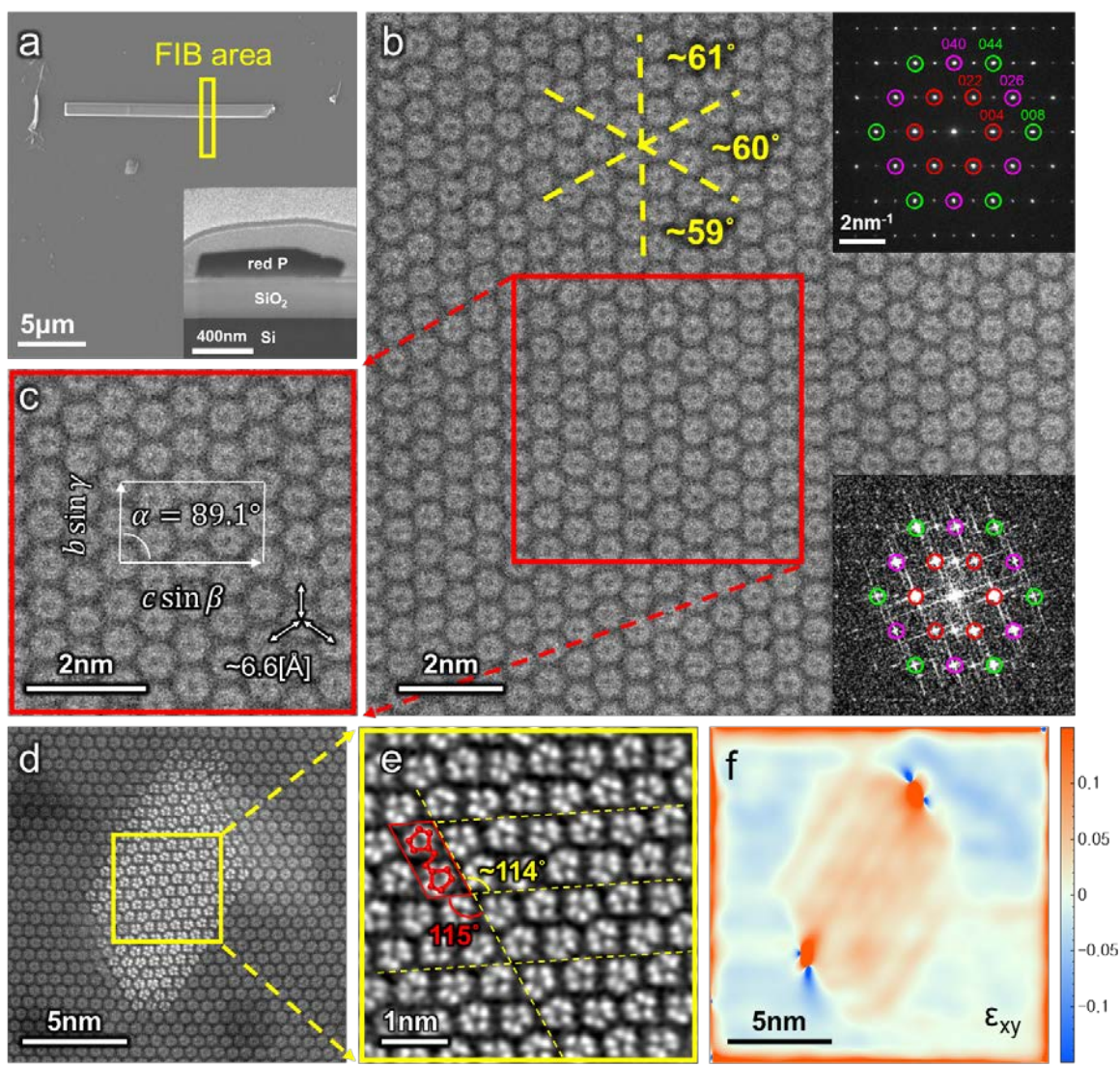

**Figure 3. Tubular structure analysis from cross-sectional HAADF-STEM imaging**. (a) SEM image of an RP nanowire used to fabricate a cross-sectional TEM sample. The inset shows the cross-sectional SEM image of the fabricated sample. (b) High-resolution cross-sectional STEM image of type-II RP. The inset shows SAED (top right) and the fast Fourier transform (bottom right) of the image. The diffraction signals are indexed according to the double-unit cell. (c) Enlarged STEM image with an overlay of unit-cell information and inter-tube distance. (d) High-resolution cross-sectional STEM image from a sample with an ultra-thin region. (e) Magnified-STEM image from the ultra-thin region, showing the close-packing of pentagon tubes. The paired tubes observed in type-IV RP are identified and marked with tubular atomic model and red box. (f) Strain map of the panel with the $\varepsilon_{xy}$ component. The distortion of the internal packing structure is observed.

Atomic resolution high-angle annular dark-field STEM (HAADF-STEM) imaging was performed to directly elucidate the complex structure of type-II RP. First, STEM imaging was performed along the nanowire axis by preparing cross-sectional samples, as shown in Figure 3a. The cross-sectional samples were prepared from individual wires via a focused ion beam. Remarkably, STEM imaging along the cross section of the nanowire clearly displayed the close packing of tubular units, as shown in Figure 3b. The electron diffraction and fast Fourier transform (FFT) were consistent with the unit cell information identified via 3D diffraction, and each diffraction signal was indexed using the double-unit cell. Eight circular tubes reside in the unit cell along the wire axis, as shown in Figure 3c. The packing of the tubes appeared to be isotropic, and the inter-tubular distance was approximately $6.6 Å$. In addition, the packing of circular units along the nanowire axis (*w*-axis) was consistent with the nanowire morphology; the easy cleavage through the surface of the packing motif resulted in the wire morphology extending along the *w*-axis.

In the thinned regions of the cross-sectional samples, the tubular motif does not exhibit a circular shape but appears as a pentagon, as shown in Figure 3e. Under STEM imaging, type-II RP undergoes local thinning via e-beam-induced sputtering, as shown in Supporting Figure S5. The thinner regions can be located by brighter intensities under STEM imaging, possibly because of better zone alignment in the thinner regions (Figure 3d). Interestingly, the pairing of pentagons (marked with red boxes) with hints of inter-pentagon connections was identified, which resembles the basic building block of type-IV RP. Compared to the thicker regions, the thin regions also display an overall distortion in the packing of the tubular structure such that the packing of pentagons resembles that observed in type-IV RP. Geometric phase analysis shows that the packing of the tubules in thin regions is distorted by approximately 8% in terms of $\varepsilon_{xy}$, compared to thicker regions, as shown in Figure 3f and Supporting Figure S6. This observation indicates that the type-II structure shares a basic building motif with type-IV RP, but its structure can be regarded as a variation with a higher complexity in terms of the internal packing structure.

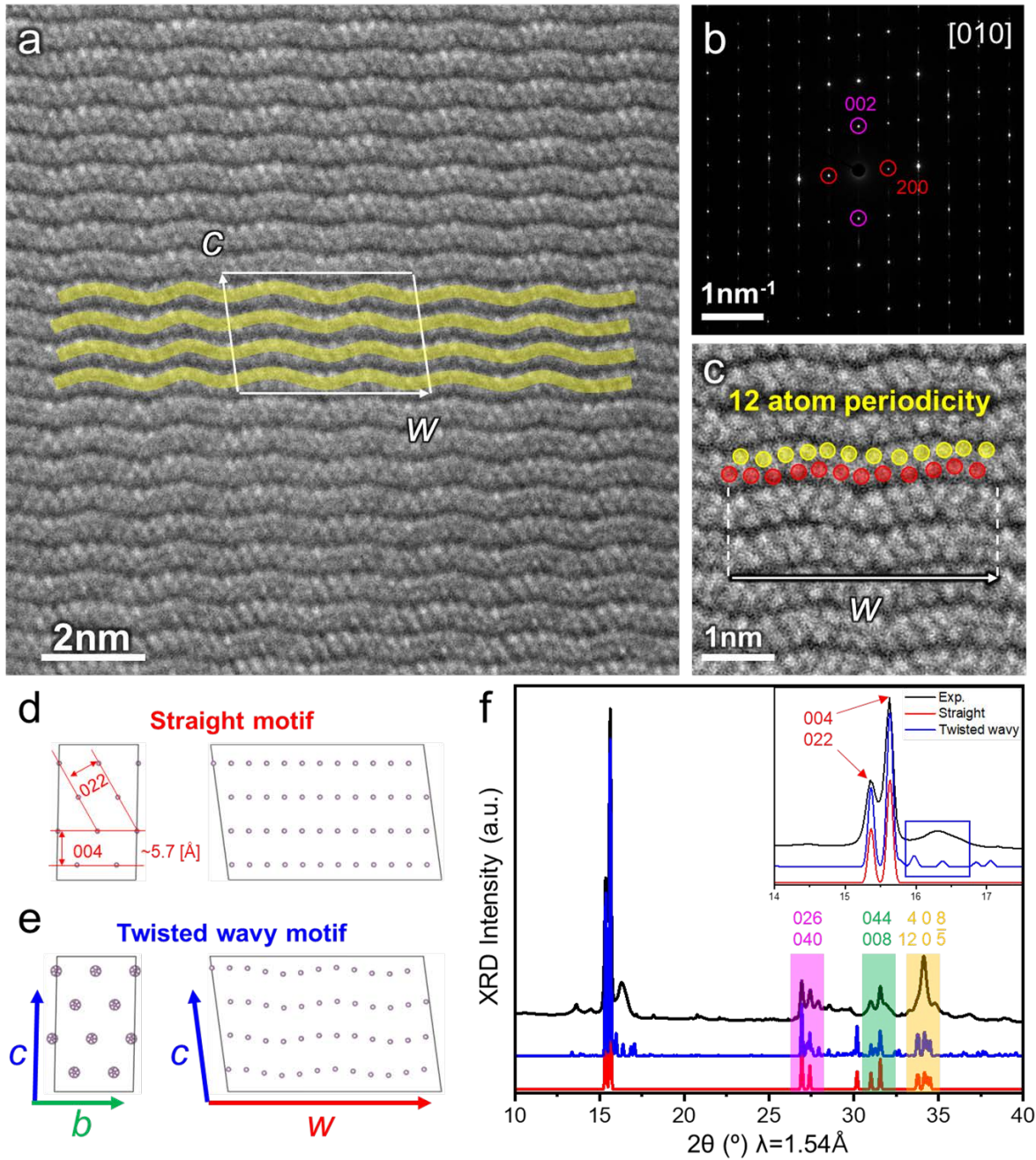

**Figure 4. Confirmation of the wavy packing structure by STEM imaging and powder XRD.** (a) Side-view of the STEM image of a type-II RP nanowire at [010] zone axis. The information of the unit-cell dimension is overlaid. (b) SAED of type-II RP from [010] zone axis. The lowest-index diffraction peaks are indexed. (c) Higher-magnification STEM image at [010] zone axis. The positions with higher intensity are marked. (d) Model of the structure with packing of straight atomic chains. The axis of atomic chains is along the *w*-axis. (e) Model of the structure with wavy packing of atomic chains. (f) Comparison of experimental and simulated XRD results based on the model of the structure. The inset shows the XRD near the strongest peaks ranging from $2\theta = 14°$ to $17.5°$.

The complexity of the packing of pentagonal tubes can be observed in a more detailed analysis of the STEM images from thin regions. As shown in Figure S7, various orientations of pentagonal pairs can be constructed by rotating the type-IV units along the tubular axis. We identified various orientations of pentagon pairs in the thin samples. These pairs often exhibit domain structures in which the pair orientations are locally well aligned. Moreover, near the boundaries between the thicker and thinner regions, various orientations of the pentagon pairs emerge. This observation strongly supports the hypothesis that type-II pentagonal pairs are present in various orientations, leading to a more complex structure.

HAADF-STEM imaging from the side view of the nanowire provided critical information on the internal structure of the type-II RP. Figure 4a-c display STEM images and SAED pattern, which were obtained from the side view and correspond to the [010] zone axis of type-II RP. The high-resolution STEM images show a distinct wavy pattern of chain structures (Figure 4a). By overlaying the unit cell identified through 3D ED onto the observed STEM image, we found that type-II RP unit cell includes four wavy chains in the *c*-axis direction. Additionally, the internal structure of a chain exhibits 12-atom periodicity along the *w*-axis inside a unit cell (Figure 4c). Our observations confirm that the basic packing motif of type-II RP consists of wavy tubular units, in contrast to the straight tubular motifs of type-IV[25] and type-V RP.[23] Combining the overall structural data elucidates the internal packing of type-II RP; the pentagonal tubular pairs are connected in a wavy pattern (instead of a straight connection) with internal rotations of the pentagons. Therefore, we term the observed structure "twisted wavy" RP.

The observed twisted wavy structure was consistent with the experimental powder XRD data. We simulated the XRD data with simplified structural models and compared them with the experimental XRD data (Figure 4d–f). The simulated XRD pattern from the packing of the simple straight atomic chain model shows an overall agreement with the experimental data, as shown in Figure 4f. In particular, the strongest peaks near $2\theta = 15.3°$ and $15.7°$ are reproduced well by the straight atomic chain model. These main peaks, indexed as 022 or 004, are associated with the interplanar distance between the tubes, as shown in Figure 4d. However, a relatively weak shoulder peak at $2\theta = 16.2°$ is absent in the straight-chain model. By introducing the wavy structure, a weak peak was reproduced (Figure 4f inset blue box), providing evidence that the building block of type-II RP is a wavy tubular structure. The twisted wavy tubular motif observed in our study shares a key aspect with structures previously suggested via theoretical calculations.[6, 12]

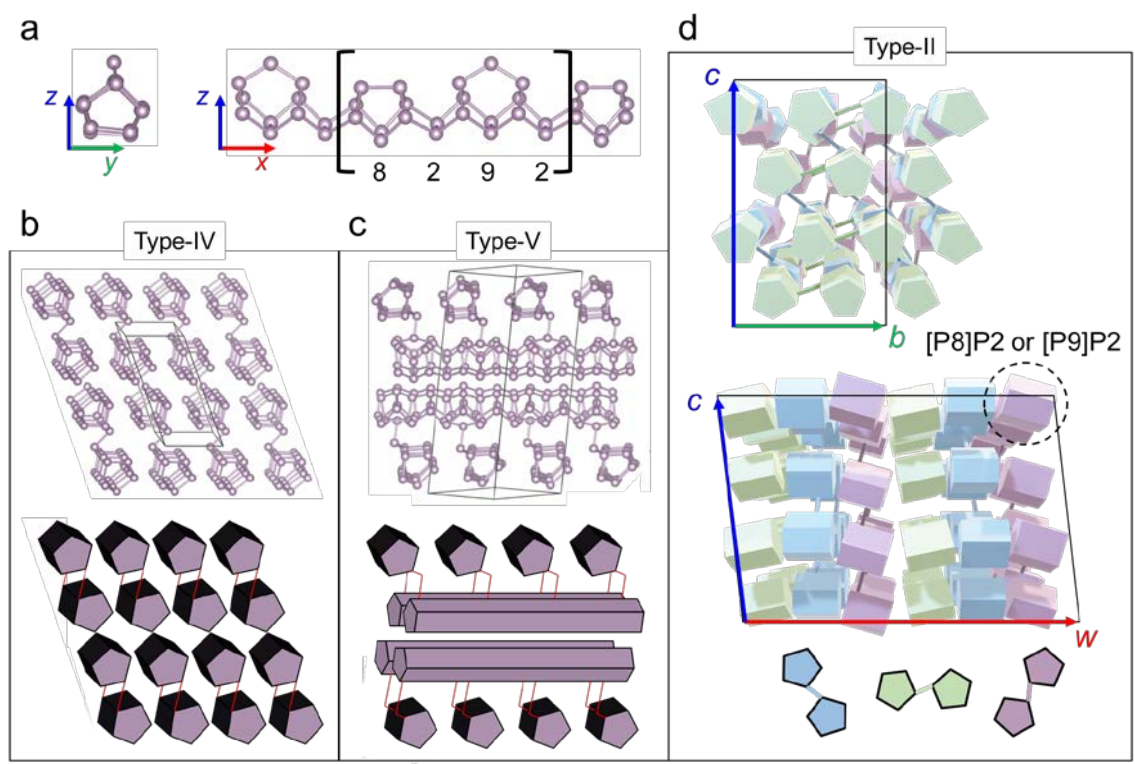

**Figure 5. Structural comparison between different crystalline phases of RP.** (a) [P8]P2[P9]P2[ tubular motif with the pentagonal cross section. P9 serves as a linker for inter-tubular connection. (b) Type-IV RP with the packing of tubular motifs along one direction. The bottom schematic shows the simplified version of the structure. (c) Type-V RP with the alternating cross-packing of tubular motifs. The bottom schematic shows the simplified version of the structure. (d) Schematic of a type-II RP structure with twisted wavy tubular motifs.

Based on the findings of this study, the comparisons of various RP polymorphs are summarized in Figure 5 and Table 1. Type-V and type-IV RP can be considered as packing structures with a straight tubular motif and a [P8]P2[P9]P2[ unit, as shown in Figure 5a–c. The main difference between the two types of RP is that type-V is an orthogonal stacked structure, whereas type-IV is a parallel packing structure of the motifs. To introduce waviness and internal twisting, the packing motif of type-II RP is likely to be more complex than [P8]P2[P9]P2[. Such candidates can be constructed based on for example [P8]P2[P9]P2[P8]P2[ unit (31 atoms) or [P9]P2[P9]P2[P8]P2[ unit (32 atoms), the variation of [P8]P2[P9]P2[. One of such candidate based on twisted wavy motif is displayed in the bottom panel of Figure 5d, in which the number of phosphorus atoms in a unit cell is approximately 250 atoms (Table 1).

**Table 1. Unit cell information of the type-II, type-IV, and type-V RP**

|  | **Type-II** (Primitive unit cell) | **Type-II** (Double unit cell) | **Type-IV** | **Type-V** |
|---|---|---|---|---|
| **Name** | Twisted wavy P | | Fibrous P | Violet P Hittorf's P |
| **System** | Triclinic | | Triclinic | Monoclinic |
| **Space group** | $P1$ or $P\bar{1}$ | | $P\bar{1}$ | $P2/n$ |
| $a$ [Å] | 18.8 | 35.5 | 12.198 | 9.210 |
| $b$ [Å] | 13.2 | | 12.986 | 9.128 |
| $c$ [Å] | 22.9 | | 7.075 | 21.893 |
| $α$ | 89.1° | | 116.99° | 90° |
| $β$ | 97.6° | | 106.31° | 97.766° |
| $γ$ | 108.8° | 88.5° | 97.91° | 90° |
| **# of atoms in the unit cell** | ~ 248* | ~ 496* | 42 | 84 |
| **Reference** | This study | | M. Ruck et al.[25] | H. Thurn et al.[23] |

* Based on [P8]P2[P9]P2[P8]P2[ tubular building blocks

**Conclusion**

This study provided an in-depth structural characterization and analysis of type-II RP for the first time. Although the structure of type-II RP consisted of a tubular packing motif packed in one direction, the tubular packing motif exhibited internal twisting and wavy patterns along the crystal wire direction. Our study clearly demonstrated that type-II RP serves as a bridge for structural conversion from the amorphous phase (type-I) to more crystalline phases. This study discovered a new variation in the building blocks of phosphorus and provided insights to clarify the complexities observed in phosphorus as well as other relevant systems.

**Methods**

**Type-II RP synthesis:** Type-II RP was synthesized using chemical vapor transport (CVT). The growth parameters such as temperatures and time was adapted from the literature (see Supporting Figure S1).[35] RP powder (100 mg, purchased from Alfa Aesar Catalog no. 010670), $I_2$ (100 mg), and a Sn-coated $SiO_2$/Si wafer were sealed using a torch in a quartz ampule in a $N_2$ environment. The ampule was placed in a furnace, and it was heated up to 600 °C and cooled (Supporting Figure S1). The ampule was opened inside a $N_2$-filled glovebox.

**Sample preparations**: TEM samples were prepared via sonication or mechanical exfoliation. An approximately 2 mm size sample was placed in a vial containing 10 mL of IPA and sonicated for 4 h. The sonicated solution was dropped onto TEM grids. The cross-sectional TEM samples were prepared using focused ion beam with a Zeiss Crossbeam 540.

**Characterizations:**

3D EDT data were collected using a JEOL microscope (JEM-2100P) with the control of EDT-Collect program at 200 kV. In a typical experimental, around 600 ED frames were recorded with the tilting angle range from around -60° to 60° using a step of 0.2°. The collected data were processed using the software EDT-Process. Atomic resolution HAADF-STEM imaging and EDS mapping were performed using a JEOL double Cs-corrected ARM-200F instrument operated at 200 kV. For HAADF-STEM imaging, we used 23 mrad convergence angle and collection semi-angles ranging from 40 to 160 mrad. The zone axis was confirmed by comparing the theoretical SAED patterns with the simulated SAED patterns. STEM images and SAED patterns were simulated by MacTempas software and SingleCrystal™ of CrystalMaker®. An open-source program, Strain++, was used to perform geometric phase analysis (GPA). Powder XRD analysis was performed using a SmartLab HR-XRD using 1.54 Å wavelength. A beam line (6D) at the Pohang Accelerator Laboratory (PAL) was used for high-resolution XRD analysis. The X-ray wavelength used in this experiment was 0.9795 Å. For comparison with normal lab-based XRD results, the synchrotron-source XRD data were converted to $2\theta$ values at a wavelength of 1.54 Å. Raman spectroscopy was performed using a Horiba LabRAM ARAMIS (532 nm laser, 50 mW, D3 filter, 1800 g/mm grating type, and 100X objective lens). Polarized Raman spectroscopy was performed using a home-built set up with a wavelength of 785 nm, power of 200 $\mu$W, and vacuum level less than $10^{-5}$ torr.

**Acknowledgements**

This study was supported by the Basic Science Research Program of the National Research Foundation of Korea (NRF-2022R1A2C4002559, NRF-2021R1C1C2006785, NRF-2022R1A2C1006740, NRF-2017R1A5A1014862, and NRF-2022R1A2C2091815) and Institute for Basic Science (IBS-R026-D1). Y. M. acknowledges support from National Science Foundation of China (22222108).

**Keywords:** Red phosphorus, Electron Microscopy, Structure Elucidation, Twisted Pentagonal Tubes, Wavy Packing Motif

**REFERENCES**

[1] D. E. Corbridge, *Phosphorus: chemistry, biochemistry and technology*, CRC press, **2013**.

[2] W. L. Roth, T. W. DeWitt, A. J. Smith, *J. Am. Chem. Soc.* **1947**, *69*, 2881-2885.

[3] M. Haeser, *J. Am. Chem. Soc.* **1994**, *116*, 6925-6926.

[4] A. J. Karttunen, M. Linnolahti, T. A. Pakkanen, *ChemPhysChem* **2008**, *9*, 2550-2558.

[5] J. M. Zaug, A. K. Soper, S. M. Clark, *Nat. Mater.* **2008**, *7*, 890-899.

[6] F. Bachhuber, J. von Appen, R. Dronskowski, P. Schmidt, T. Nilges, A. Pfitzner, R. Weihrich, *Angew. Chem. Int. Ed.* **2014**, *53*, 11629-11633.

[7] Y. Zhou, W. Kirkpatrick, V. L. Deringer, *Adv. Mater.* **2022**, *34*, 2107515.

[8] S. Böcker, M. Häser, *Z. Anorg. Allg. Chem.* **1995**, *621*, 258-286.

[9] J. Zhang, D. Zhao, D. Xiao, C. Ma, H. Du, X. Li, L. Zhang, J. Huang, H. Huang, C.-L. Jia, D. Tománek, C. Niu, *Angew. Chem.* **2017**, *129*, 1876-1880.

[10] M. Hart, E. R. White, J. Chen, C. M. Mcgilvery, C. J. Pickard, A. Michaelides, A. Sella, M. S. P. Shaffer, C. G. Salzmann, *Angew. Chem.* **2017**, *129*, 8256-8260.

[11] G. Sansone, L. Maschio, A. J. Karttunen, *Chem. Eur. J.* **2017**, *23*, 15884-15888.

[12] V. L. Deringer, C. J. Pickard, D. M. Proserpio, *Angew. Chem. Int. Ed.* **2020**, *59*, 15880-15885.

[13] Z. Zhu, D. Tománek, *Phys. Rev. Lett.* **2014**, *112*, 176802.

[14] D. V. Rybkovskiy, V. O. Koroteev, A. Impellizzeri, A. A. Vorfolomeeva, E. Y. Gerasimov, A. V. Okotrub, A. Chuvilin, L. G. Bulusheva, C. P. Ewels, *ACS Nano* **2022**, *16*, 6002-6012.

[15] L. Zhang, H. Huang, B. Zhang, M. Gu, D. Zhao, X. Zhao, L. Li, J. Zhou, K. Wu, Y. Cheng, J. Zhang, *Angew. Chem. Int. Ed.* **2020**, *59*, 1074-1080.

[16] G. Cicirello, M. Wang, Q. P. Sam, J. L. Hart, N. L. Williams, H. Yin, J. J. Cha, J. Wang, *J. Am. Chem. Soc.* **2023**, *145*, 8218-8230.

[17] A. Pfitzner, M. F. Bräu, J. Zweck, G. Brunklaus, H. Eckert, *Angew. Chem. Int. Ed.* **2004**, *43*, 4228-4231.

[18] L. Li, Y. Yu, G. J. Ye, Q. Ge, X. Ou, H. Wu, D. Feng, X. H. Chen, Y. Zhang, *Nat. Nanotechnol.* **2014**, *9*, 372-377.

[19] G. Schusteritsch, M. Uhrin, C. J. Pickard, *Nano Lett.* **2016**, *16*, 2975-2980.

[20]   Z. Yang, J. Hao, S. Yuan, S. Lin, H. M. Yau, J. Dai, S. P. Lau, *Adv. Mater.* **2015**, *27*, 3748-3754.

[21]   P. E. M. Amaral, G. P. Nieman, G. R. Schwenk, H. Jing, R. Zhang, E. B. Cerkez, D. Strongin, H. F. Ji, *Angew. Chem. Int. Ed.* **2019**, *58*, 6766-6771.

[22]   L. Du, Y. Zhao, L. Wu, X. Hu, L. Yao, Y. Wang, X. Bai, Y. Dai, J. Qiao, M. G. Uddin, X. Li, J. Lahtinen, X. Bai, G. Zhang, W. Ji, Z. Sun, *Nat. Commun.* **2021**, *12*, 4822.

[23]   H. Thurn, H. Kerbs, *Angew. Chem. Int. Ed.* **1966**, *5*, 1047-1048.

[24]   A. G. Ricciardulli, Y. Wang, S. Yang, P. Samorì, *J. Am. Chem. Soc.* **2022**, *144*, 3660-3666.

[25]   M. Ruck, D. Hoppe, B. Wahl, P. Simon, Y. Wang, G. Seifert, *Angew. Chem. Int. Ed.* **2005**, *44*, 7616-7619.

[26]   R. A. L. Winchester, M. Whitby, M. S. P. Shaffer, *Angew. Chem. Int. Ed.* **2009**, *48*, 3616-3621.

[27]   Z. Shen, Z. Hu, W. Wang, S.-F. Lee, D. K. L. Chan, Y. Li, T. Gu, J. C. Yu, *Nanoscale* **2014**, *6*, 14163-14167.

[28]   J. B. Smith, Daniel Hagaman, David Diguiseppi, Reinhard Schweitzer-Stenner, Hai-Feng Ji, *Angew. Chem. Int. Ed.* **2016**, *55*, 11829-11833.

[29]   Q. Liu, X. Zhang, J. Wang, Y. Zhang, S. Bian, Z. Cheng, N. Kang, H. Huang, S. Gu, Y. Wang, D. Liu, P. K. Chu, X. F. Yu, *Angew. Chem. Int. Ed.* **2020**, *59*, 14383-14387.

[30]   Y. Zhu, J. Ren, X. Zhang, D. Yang, *Nanoscale* **2020**, *12*, 13297-13310.

[31]   L. Zhang, H. Huang, Z. Lv, L. Li, M. Gu, X. Zhao, B. Zhang, Y. Cheng, J. Zhang, *ACS Applied Electronic Materials* **2021**, *3*, 1043-1049.

[32]   J. Li, J. Sun, *Acc. Chem. Res.* **2017**, *50*, 2737-2745.

[33]   L. Meshi, S. Samuha, *Adv. Mater.* **2018**, *30*, 1706704.

[34]   P. Brázda, L. Palatinus, M. Babor, *Science* **2019**, *364*, 667-669.

[35]   L. Zhang, M. Gu, L. Li, X. Zhao, C. Fu, T. Liu, X. Xu, Y. Cheng, J. Zhang, *Chem. Mater.* **2020**, *32*, 7363-7369.